\documentstyle[prl,aps,epsf,floats]{revtex}

\begin{document}
\draft

\twocolumn[\hsize\textwidth\columnwidth\hsize\csname @twocolumnfalse\endcsname
\title{
Effect of impurities on transport through organic self-assembled molecular films from first 
principles 
}
\author{B. Larade and A.M. Bratkovsky }
 
\address{
Hewlett-Packard Laboratories, 1501 Page Mill Road, Palo Alto, 
California 94304
}
\date{July 31, 2004}
\maketitle
\begin{abstract}
We calculate electron transport through molecular monolayers of
saturated alkanes  with point defects from first principles. 
Single defects (incorporated Au ions, kinks, 
dangling bonds) produce deep localized levels in the gap between
occupied and unoccupied molecular levels. Single defects 
produce steps on the I-V curve, whereas pairs of
(unlike and like) defects give negative differential resistance peaks. 
The results are discussed  
in relation to the observed unusual transport behavior of organic monolayers
and compared with transport through conjugated polythiophenes.
\end{abstract}
\narrowtext

\vskip2pc] Recently, an interest towards the development and application of
Langmuir-Blodgett films and self-assembled monolayers (SAM)\ has risen
dramatically, which is in part related to widespread efforts in the area of
molecular-size devices and sensors. At the same time, the fundamental
understanding of these films is assessed as ``very limited'' \cite{Ulman95}.
Further, it has been concluded some decades ago that the conduction through
absorbed \cite{poly78} and Langmuir-Blodgett \cite{tredgold81} monolayers of
fatty acids (CH$_{2})_{n}$ (we shall customarily refer to them as C$n$)
is associated with defects. In particular, Polymeropoulos and Sagiv have
studied a variety of absorbed monolayers from C7 to C23 on Al/Al$_{2}$O$_{3}$
substrates and found that the exponential dependence on the length of the
molecular chains is only observed below the liquid nitrogen temperature of
77K, and no discernible length dependence was observed at higher
temperatures \cite{poly78}. The current varied strongly with the temperature
for LB films on Al/Al$_{2}$O$_{3}$ substrates in He atmosphere (believed to
hinder the Al$_{2}$O$_{3}$ growth)\cite{tredgold81}, which is not compatible
with elastic tunneling. The 100-fold increase of resistance of the annealed
films was attributed to the ``annealing'' of \ ``defects'' with time. An
absence of direct tunneling through self-assembled monolayers of C12-C18 has
been reported by Boulas et al. \cite{boulas96}. On the other hand, an
exponential dependence of monolayer resistance on the chain length $L$, $%
R_{\sigma }\propto \exp (\beta _{\sigma }L),$ \cite{sakaguchi01,reed03} and
no temperature dependence of the conductance in C8-C16 molecules was
observed over the temperatures $T=80-300$K, a signature of the elastic
tunneling \cite{reed03}.

The electrons in alkane molecules are tightly bound to the C atoms by $%
\sigma -$bonds, and the band gap (between the highest occupied molecular
orbital, HOMO, and lowest unoccupied molecular orbital, LUMO) is large, $%
\sim 9-10$eV \cite{boulas96}, Fig.~1a. In conjugated systems with $\pi -$%
electrons the molecular orbitals are extended, and the HOMO-LUMO\ gap is
correspondingly smaller, as in e.g. polythiophenes, where the resistance was
also found to scale exponentially with the length of the chain, $R_{\pi
}\propto \exp (\beta _{\pi }L),$ with $\beta _{\pi }=0.35$\AA $^{-1}$
instead of $\beta _{\sigma }=1.08$\AA $^{-1}$\cite{sakaguchi01}. In 
contrast with the temperature-independent tunneling results for SAMs \cite
{reed03}, recent extensive studies of electron transport through 2.8 nm
thick eicosanoic acid (C20) LB\ monolayers at temperatures 2K-300K have
established that the current is practically temperature independent below $%
T<60$K, but very strongly temperature dependent at higher temperatures $%
T=60-300$K\cite{DunTdep04}. 
\begin{figure}[t]
\epsfxsize=3.in 
\epsffile{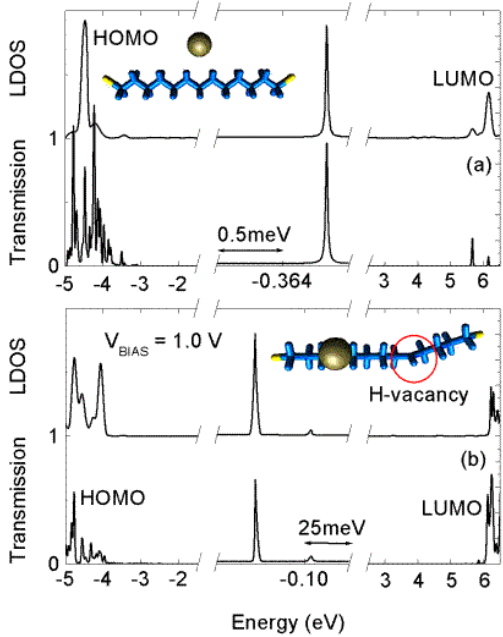}
\caption{ Local density of states and transmission as a function of energy
for (a) C13 with Au impurity and (b) C13 with Au impurity and H vacancy
(dangling bond). Middle sections show closeups of the resonant peaks due to
deep defect levels with respect to the HOMO and LUMO molecular states. The
HOMO-LUMO gap is about 10 eV. }
\label{fig:FigTR}
\end{figure}

A large amount of effort went into characterizing the organic thin films and
possible defects there \cite{Ulman95,allara00,allaranoble04}. It has been
found that the electrode material, like gold, gets into the body of the
film, leading to the possibility of metal ions existing in the film as
single impurities and clusters. Electronic states on these impurity ions are
available for the resonant tunneling of carriers in very thin films (or
hopping in thicker films, a crossover between the regimes depending on the
thickness). Depending on the density of the impurity states, with increasing
film thickness the tunneling will be assisted by impurity ``chains'', with
an increasing number of equidistant impurities \cite{pollak73}. One-impurity
channels produce steps on the I-V curve but no temperature dependence,
whereas the inelastic tunneling through pairs of impurities at low
temperatures defines the temperature dependence of the film conductance, $%
G(T)\propto T^{4/3},$ and the voltage dependence of current $I(V)\propto
V^{7/3}$\cite{GlaMat88}. This is a generic behavior observed in inorganic
systems with defects\cite{Mac95,Mood98}. The tunneling may be accompanied by
interaction with vibrons on the molecule, causing step-like features on the
I-V curve \cite{zhitvibr,AB03vibr}. During processing, especially top
electrode deposition, small clusters of the electrode material may form in
the organic film, causing Coulomb blockade, which also can show up as steps
on the I-V curve. It has long been known that a strong applied field can
cause localized damage to thin films, presumably due to electromigration and
the formation of conducting filaments\cite{LBfila8671}. Recent spatial
mapping of the conductance in LB\ monolayers of fatic acids with the use of
conducting AFM\ has revealed damage areas 30-100nm in diameter appearing
after a ``soft'' electrical breakdown, sometimes accompanied by a strong
temperature dependence of the conductance\cite{Jennie04}. A crossover from
tunneling at low temperatures to an activation-like dependence at higher
temperatures is expected for electron transport through organic molecular
films. There are recent reports about such a crossover in individual
molecules like the 2nm long Tour wire with a small activation energy $%
E_{a}\approx 130$meV \cite{allaraBJTdep04}. Our present results suggest that
this may be a result of interplay between the drastic renormalization of the
electronic structure of the molecule in contact with electrodes, and
disorder in the film, illustrated by Fig. 2 (inset).

We report the ab-initio calculations of point-defect assisted tunneling
through alkanedithiols S(CH$_{2})_{n}$S and thiophene T3 (three rings SC$%
_{4})$ self-assembled on gold electrodes. The length of the alkane chain was
in the range $n=9-15$. We have studied single and double defects in the
film: (i) single Au impurity, Figs. 1a, 2a, (ii) Au impurity and H vacancy
(dangling bond) on the chain, Fig. 1b, 2c, (iii) a pair of Au impurities,
Fig. 2b, (iv) Au and a ``kink'' on the chain (one C=C bond instead of a C-C
bond). Single defect states result in steps on the associated I-V curve,
whereas molecules in the presence of two defects generally exhibit a
negative differential resistance (NDR). Both types of behavior are generic
and may be relevant to some observed unusual transport characteristics of
SAMs and LB\ films \cite
{poly78,tredgold81,DunTdep04,Jennie04,allaraBJTdep04,Zhit04}.

We have used an ab-initio approach that combines the Keldysh non-equilibrium
Green's function (NEGF) method with self-consistent pseudopotential-based
real space density functional theory (DFT) for systems with {\em open}{\it \ 
}boundary conditions provided by semi-infinite electrodes under external
bias voltage \cite{Guo01,BLABnaph}. To construct the system Hamiltonian, we
use self-consistent DFT within the local-density approximation. The atomic
cores are defined using standard norm-conserving nonlocal pseudopotentials 
\cite{martins91}, and we expand the Kohn-Sham wave functions in a fireball 
\cite{sankey89} s-, p-, d- real-space atomic orbital basis\cite{Guo01}. We
have checked that the results are reliable with respect to the size of the
basis set. The Greens function is determined by direct matrix inversion.
External bias is incorporated into the Hartree potential, and in this way
the nonequilibrium charge density is iterated to self-consistency with
imposed global charge neutrality, after which the current-voltage (I-V)
characteristics have been computed. All present structures have been relaxed
with the Gaussian98 code prior to transport calculations\cite{g98}. The
conductance of the system at a given energy is found from 
\begin{equation}
g(E,V)=\frac{4}{q}{\rm Tr}\left[ \left( 
\mathop{\rm Im}%
\Sigma _{l,l}\right) G_{l,r}^{R}\left( 
\mathop{\rm Im}%
\Sigma _{r,r}\right) G_{r,l}^{A}\right] ,  \label{eq:cond}
\end{equation}
where $G^{R(A)}$ is the retarded (advanced)\ Green's function, $\Sigma $ is
the self-energy part connecting left ($l$) and right ($r$) electrodes \cite
{Guo01}, and the current is found as 
\begin{equation}
I=\frac{2q^{2}}{h}\int dE[f(E-qV/2)-f(E+qV/2)]g(E,V).  \label{eq:curr}
\end{equation}
\begin{figure}[t]
\epsfxsize=2.8in \epsffile{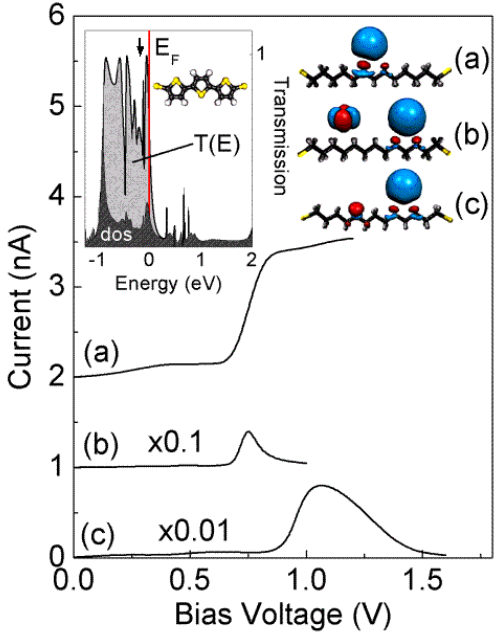}
\caption{ Current-voltage characteristics of an alkane chain C13 with (a)
single Au impurity (6s-state), (b) two Au impurities (5d and 6s-states on
left and right ions, respectively), and (c) Au impurity and H vacancy
(dangling bond). Double defects produce the negative differential resistance
peaks (b) and (c). Inset shows the density of states, transmission, and
stick model for polythiophene T3. There is significant transmission at the
Fermi level, suggesting an ohmic I-V characteristic for T3 connected to gold
electrodes. \ Disorder in the film may localize states close to the Fermi
level (schematically marked by arrow), which may assist in hole hopping
transport with an apparently very low activation energy (0.01-0.1eV), as is
observed.}
\label{fig:fig2iv}
\end{figure}
The equilibrium position of an Au impurity is about 3\AA\ away from the
alkane chain, which is a typical Van-der-Waals distance. Note that the
conductance of alkanes is dominated by tunneling through the wide HOMO-LUMO\
gap (see below) and, therefore, is not sensitive to fine details of
molecule-electrode coupling, unlike in the case of conjugated molecules,
where the sensitivity to the geometry may be very high\cite{BKgeom03}. As
the density maps show (Fig. 2), there is an appreciable hybridization
between the s- and d-states of Au and the sp- states of the carbohydride
chain. Furthermore, the Au ion produces a Coulomb center trapping a 6s
electron state at an energy $\epsilon _{i}=-0.35$eV with respect to the
Fermi level, almost in the middle of the HOMO-LUMO $\sim 10$eV\ gap in C$_{n}
$ (DFT gaps are usually smaller than in experiment). The tunneling
evanescent resonant state is a superposition of the HOMO and LUMO\ molecular
orbitals. Those orbitals have a very complex spatial structure, reflected in
an asymmetric line shape for the transmission. Since the impurity levels are
very deep, they may be understood within the model of \ ``short-range
impurity potential'' \cite{LarMat87}. Indeed, the impurity wave function 
{\em outside} of the narrow well can be fairly approximated as 
\begin{equation}
\varphi (r)=C \exp(-\kappa r)/r,  \label{eq:fii}
\end{equation}
where $C=\sqrt{\kappa/2\pi}$, $\kappa $ is the inverse radius of the state, $\hbar ^{2}\kappa
^{2}/2m^{\ast }=E_{i},$ where $E_{i}=\Delta -\epsilon _{i}$ is the depth of
the impurity level with respect to the LUMO, and $\Delta $=LUMO$-F$ is the
distance between the LUMO\ and the Fermi level of gold and, consequently,
the radius of the impurity state $1/\kappa $ is small. The energy distance
is $\Delta \approx 4.8$eV in alkane chains (CH$_{2})_{n}$\cite{boulas96}{\bf %
\ (}$\approx 5$eV from DFT calculations), and $m^{\ast }\sim 0.4$ is the
effective electron tunneling mass in alkanes\cite{reed03}. For one impurity
in a rectangular tunnel barrier \cite{LarMat87} we obtain the Breit-Wigner
form of conductance $g(E)\;$%
\begin{equation}
g(E)=\frac{4q^{2}}{\pi \hbar }\frac{\Gamma _{L}\Gamma _{R}}{\left(
E-\epsilon _{i}\right) ^{2}+\left( \Gamma _{L}+\Gamma _{R}\right) ^{2}/4},
\label{eq:TEbw}
\end{equation}
where $\Gamma _{L(R)}/\hbar $ is the tunneling rate from impurity to the
left (right)\ electrode, $\Gamma _{L(R)}\sim t^{2}/D=\Gamma _{0}e^{-2\kappa
L_{1(2)}},$ where $t$ is the tunneling amplitude from impurity\ to the
electrode, $D$ is the electron band width in the electrodes. Using this
model, we may estimate for an Au impurity in C13 ($L=10.9$\AA ) the width $%
\Gamma _{L}=\Gamma _{R}=1.2\times 10^{-6}$, which is within an order of
magnitude compared with the value $1.85\times 10^{-5}$eV calculated from (%
\ref{eq:cond}). The transmission is maximal and equals unity when $%
E=\epsilon _{i}$ and $\Gamma _{L}=\Gamma _{R},$ which corresponds to a
symmetrical position for the impurity with respect to the electrodes. The
electronic structure of the alkane backbone, through which the electron
tunnels to an electrode, shows up in the asymmetric lineshape, which is
substantially non-Lorentzian, Fig. 1. The current remains small until the
bias has aligned the impurity level with the Fermi level of the electrodes,
resulting in a step in the current , $I_{1}\approx \frac{2q}{\hbar }\Gamma
_{0}e^{-\kappa L}$ (Fig.2a)$.$ This step can be observed only when the
impurity level is not very far from the Fermi level $F,$ such that biasing
the contact can produce alignment before a breakdown of the device may occur.

The most interesting situations that we have found relate to the {\em pairs}
of point defects in the film. Indeed, some impurity levels on two sites may
move in and out of resonance, see Eq.(\ref{eq:levels}), and this obviously
will lead to peaks on the I-V\ curve, or, equivalently, to Negative
Differential Resistance (NDR). If the concentration of defects is $c\ll 1$,
the relative number of configurations with pairs of impurities will be
small, $\propto c^{2}.$ However, they give an exponentially larger
contribution to the current. Indeed, the optimal position of two impurities
is symmetrical, a distance $L/2$ apart, with current $I_{2}\propto
e^{-\kappa L/2}.$ The conductance of a {\em two-impurity chain} is \cite
{LarMat87} 
\begin{equation}
g_{12}(E)=\frac{4q^{2}}{\pi \hbar }\frac{\Gamma _{L}\Gamma _{R}t_{12}^{2}}{%
\left| (E-\epsilon _{1}+i\Gamma _{L})(E-\epsilon _{2}+i\Gamma
_{R})-t_{12}^{2}\right| ^{2}}.  \label{eq:g12}
\end{equation}
For a pair of impurities with slightly differing energies $t_{12}=\frac{%
2(E_{1}+E_{2})e^{-\kappa r_{12}}}{\kappa r_{12}},$where $r_{12}$ is the
distance between them. The interpretation of the two-impurity channel
conductance (\ref{eq:g12})\ is fairly straightforward:\ if there were no
coupling to the electrodes, i.e. $\Gamma _{L}=\Gamma _{R}=0,$ the poles of $%
g_{12}$ would coincide with the bonding and antibonding levels of the
two-impurity ``molecule''. The coupling to the electrodes gives them a
finite width and produces, generally, two peaks in conductance, whose
relative positions in energy change with the bias. The same consideration is
valid for longer chains too, and gives an intuitive picture of the formation
of the impurity ``band''\ of states. The maximal conductivity $%
g_{12}=q^{2}/\pi \hbar $ occurs when $\epsilon _{1}=\epsilon _{2},$ $\Gamma
_{L}\Gamma _{R}=t_{12}^{2}=\Gamma _{2}^{2},$ where $\Gamma _{2}$ is the
width of the two-impurity resonance, and it corresponds to the symmetrical
position of the impurities along the normal to the contacts separated by a
distance equal to half of the molecule length, $r_{12}=L/2.$ Under bias
voltage the impurity levels shift as 
\begin{equation}
\epsilon _{i}=\epsilon _{i0}+qVz_{i}/L,  \label{eq:levels}
\end{equation}
where $z_{i}$ are the positions of the impurity atoms counted from the
center of the molecule. Due to disorder in the film, under bias voltage the
levels will be moving in and out of resonance, thus producing {\em NDR\ peaks%
} on the I-V curve. The most pronounced negative differential resistance is
presented by a gold impurity next to a C$n$ chain with an H-vacancy on
one site, Fig.~1b (the defect corresponds to a dangling bond). The defects
result in two resonant peaks in transmission, Fig.~1b. Surprisingly, the H\
vacancy (dangling bond) has an energy very close to the electrode Fermi
level $F$, with $\epsilon _{i}=-0.1$eV (Fig.~1b, right peak in the middle).\
The relative positions of the resonant peaks move with an external bias and
have crossed at 1.2V, producing a pronounced NDR\ peak in the I-V curve,
Fig.~2c. No NDR\ peak is seen in the case of an Au impurity and a kink C=C\
on the chain because the energy of the kink level is far from that of the Au
6s impurity level. The calculated values of the peak current through the
molecules were large: $I_{p}\approx 90$ nA/molecule for an Au impurity with
H vacancy, and $\approx 5$ nA/molecule for double Au impurities. We have
observed a new mechanism for the NDR peak in a situation with two Au
impurities in the film. Namely, Au ions produce two sets of deep impurity
levels in C$n$ films, one stemming from the 6s orbital, another from the
5d shell, as clearly seen in Fig.~2b (inset). The 5d-states are separated in
energy from 6s, so that now the tunneling through s-d pairs of states is
allowed in addition to s-s tunneling. Since the 5d-states are at a lower
energy than the s-state, the d- and s-states on different Au ions will be
aligned at a certain bias. Due to the different angular character of those
orbitals, the tunneling between the s-state on the first impurity and a
d-state on another impurity will be described by the hopping integral
analogous to the Slater-Koster $sd\sigma $ integral. The peak current in
that case is smaller than for the pair Au-H vacancy, where the overlap is of 
$ss\sigma $ type (cf. Figs.~2b and 2c). Thiophene molecules behave very
differently since the $\pi -$states there are conjugated and, consequently,
the HOMO-LUMO\ gap is much narrower, just below 2 eV. The tail of the HOMO\
state in the T3 molecule (with three rings) has a significant presence near
the electrode Fermi level, resulting in a practically ``metallic'' density
of states and hence ohmic I-V\ characteristic. This behavior is quite robust
and is in apparent disagreement with experiment, where the tunneling has been
observed\cite{sakaguchi01}. However, in actual thiophene devices the contact
between the molecule and electrodes is obviously very poor, and it may lead
to unusual current paths and temperature dependence\cite{Zhit04}.

We have presented the first parameter-free DFT calculations of a class of
organic molecular chains incorporating single or double point defects. The
results suggest that the present generic defects produce deep impurity
levels in the film and cause a resonant tunneling of electrons through the
film, strongly dependent on the type of defects. Thus, a missing hydrogen
produces a level (dangling bond) with an energy very close to the Fermi
level of the gold electrodes $F.$ In the case of a single impurity, it
produces steps on the I-V curve when one electrode's Fermi level aligns with
the impurity level under a certain bias voltage. 

The two-defect case is much
richer, since in this case we generally see a formation of the
negative-differential resistance peaks. We found that the Au atom together
with the hydrogen vacancy (dangling bond) produces the most pronounced NDR
peak at a bias of 1.2V in C13. Other pairs of defects do not produce such
spectacular NDR peaks. A short range impurity potential model reproduces the
data very well, although the actual lineshape is different. 

There is a
remaining question of what may cause the strong temperature dependence of
conductance in ``simple'' organic films like [CH$_{2}]_{n}$. The
activation-like conductance $\propto \exp (-E_{a}/T)$ has been reported with
a small activation energy $E_{a}\sim 100-200$meV in alkanes \cite
{DunTdep04,allaraBJTdep04} and even smaller, 10-100 meV, in polythiophenes 
\cite{Zhit04}. This is much smaller than the value calculated here for
alkanes and expected from electrical and optical measurements on C$n$
molecules, $E_{a}\sim \Delta \sim 4$eV \cite{boulas96}, which correspond
nicely to the present results. In conjugated systems, however, there may be
rather natural explanation of small activation energies. Indeed, the HOMO in
T3 polythiophene on gold is dramatically broadened, shifted to higher
energies and has a considerable weight at the Fermi level. The upward shift
of the HOMO is just a consequence of the work function difference between
gold and the molecule. In the presence of (inevitable) disorder in the film
some of the electronic states on the molecules will be localized in the
vicinity of $E_{F}$. Those states will assist the thermally activated
hopping of holes within a range of small activation energies $\lesssim 0.1$%
~eV. Similar behavior is expected for Tour wires\cite{allaraBJTdep04}, where 
$E_{F}-{\rm HOMO}\sim 1$~eV\cite{Guo01}(c), if the electrode-molecule
contact is poor, as is expected. Metallic protrusions (filaments), emerging
due to electromigration in a very strong electric field, and metallic,
hydroxyl, etc. inclusions\cite{LBfila8671,Jennie04} may result in a much
smaller tunneling distance $d$ for the carriers and the image charge
lowering of the barrier. The image charge lowering of the barrier in a gap
of width $d$ is $\Delta U=q^{2}\ln 4/(\epsilon d),$ meaning that a decrease
of about 3.5eV may only happen in an unrealistically narrow gap $d=2-3$\AA %
~in a film with dielectric constant $\epsilon =2.5,$ but it will add to the
barrier lowering. More detailed characterization and theoretical studies
along these lines may help to resolve this very unusual behavior. We note
that such a mechanism cannot explain the crossover with temperature from
tunneling to hopping reported for single molecular measurements, which has
to be a property of the device, not a single molecule \cite{allaraBJTdep04}.
The work has been supported by DARPA.

\end{document}